\documentclass[
 aps, pra,
 amsmath,amssymb,
 11pt,
 final,
tightenlines,
 twoside,
 onecolumn,
 nofloats,
nofootinbib,
 superscriptaddress,
showkeys,
showkeywords,
 ]
{revtex4-2}

\usepackage[T2A]{fontenc}
\usepackage[utf8]{inputenc}
\usepackage[russian,english]{babel}
\usepackage{graphicx}% Include figure files
\usepackage{dcolumn}% Align table columns on decimal point
\usepackage{bm}% bold math

\usepackage{ulem}

\input{maik.rty}

\setcitestyle{authoryear,round}
\def\squareforqed{\hbox{\rlap{$\sqcap$}$\sqcup$}}

\def\sq{\ifmmode\squareforqed\else{\unskip\nobreak\hfil
\penalty50\hskip1em\null\nobreak\hfil\squareforqed
\parfillskip=0pt\finalhyphendemerits=0\endgraf}\fi}

\def\utw{\smash{\rlap{\lower5pt\hbox{$\sim$}}}}

\def\udtw{\smash{\rlap{\lower6pt\hbox{$\approx$}}}}

\def\diameter{{\ifmmode\mathchoice
{\ooalign{\hfil\hbox{$\displaystyle/$}\hfil\crcr
{\hbox{$\displaystyle\mathchar"20D$}}}}
{\ooalign{\hfil\hbox{$\textstyle/$}\hfil\crcr
{\hbox{$\textstyle\mathchar"20D$}}}}
{\ooalign{\hfil\hbox{$\scriptstyle/$}\hfil\crcr
{\hbox{$\scriptstyle\mathchar"20D$}}}}
{\ooalign{\hfil\hbox{$\scriptscriptstyle/$}\hfil\crcr
{\hbox{$\scriptscriptstyle\mathchar"20D$}}}}
\else{\ooalign{\hfil/\hfil\crcr\mathhexbox20D}}%
\fi}}

\def\be{\begin{equation}}
\def\ee{\end{equation}}
\def\ba{\begin{eqnarray}}
\def\ea{\end{eqnarray}}

\def\msun{M_\odot}

\def\ltsima{$\; \buildrel < \over \sim \;$}
\def\simlt{\lower.5ex\hbox{\ltsima}}
\def\gtsima{$\; \buildrel > \over \sim \;$}
\def\simgt{\lower.5ex\hbox{\gtsima}}

\usepackage[dvipsnames]{color}
\begin{document}

\selectlanguage{english}

%\keywords{shock waves -- supernova remnants -- dust}

%\ydk{}
%\titlerunning{}
%\authorrunning{}
%\toctitle{}
%\tocauthor{}

\title{Large dust grains escaping from a supernova remnant}

\author{\firstname{E.~O.}~\surname{Vasiliev}}
 \email{eugstar@mail.ru}
 \affiliation{Lebedev Physical Institute, Russian Academy of Sciences, Moscow, 117997 Russia}

\begin{abstract}
Dust grains produced in the supernova (SN) ejecta at early ages are partially destroyed in hot gas of the SN remnant. Large grains owing to their inertia can overtake the forward shock front of the remnant and move through the ambient interstellar medium. Using one-dimensional gas dynamics including the dust destruction we study the transport of dust grains inside the SN remnant and the ambient medium. Here we estimate the surviving and escaping mass fractions depending on the cut-off (maximum) size in a power-law distribution of grain sizes. We found that these fractions follow closely each other at late age of the SN remnant, i.e. almost all grains, which have survived in the remnant, leave it for the ambient interstellar medium. At the age of 30~kyr -- the onset of the radiative phase with density $n\sim 1$~cm$^{-3}$ with solar metallicity -- the escaping/surviving mass fractions for the cut-off size of several microns reach about $0.7-0.8$. The major part of these grains have size larger than $1~\mu$m and initially located at distances$\simgt 0.2$ of the ejecta core radius. Later dust grains at velocity of several hundreds km/s escaping the remnant continue to be destroyed due to kinetic sputtering in the interstellar medium. Grains with sizes of several microns are destroyed marginally and therefore will survive in the interstellar gas. After decelerating of grains in the medium the surviving mass fraction of grains produced in the SN remnant, which evolves in an ambient gas with $n$ from 1 to 100~cm$^{-3}$, is from $\sim 0.5$ to $\sim 0.3$, respectively, for the cut-off grain size $a_{cut} = 10~\mu$m. A decrease of the cut-off size in the distribution results in lower surviving mass fraction.
\end{abstract}

\maketitle

%----------------------- Section 1 -------------------------------
\section{INTRODUCTION}\label{sec:intro}

Dust is produced in ejecta of Type II supernova (SN) at early ages \citep{Kozasa1989,Kozasa2009,Todini2001,Nozawa2003,Nozawa2006,Nozawa2007,Sarangi2015,Sluder2018,
Cherchneff2009,Cherchneff2010,Cherchneff2011,Cherchneff2013,Dwek2016}. Afterwards dust grains are destroyed by sputtering in hot gas heated by both forward and reverse shocks \citep{Scalo1977,Shull1977,Tielens1987,Dwek1996,Jones1996,Janev2001}. Mostly the dust destruction efficiency  depends significantly on the passage of grains through the reverse shock \citep{Nath2008,Silvia2010,Silvia2012,Biscaro2016,Kirchschlager2019,Kirchschlager2020,Shchekinov2025}. 

\citet{Slavin2020} were apparently the first, who found that grains produced in the SN ejecta can overtake the forward shock. They showed that grains larger than tenths of a micron can survive in hostile conditions of the remnant and escape into the interstellar medium (ISM): 
about $(10-20)$\% of silicate grains survive, when the initial size distribution contains grains up to $\sim 0.5~\mu$m, and $(30-50)$\% of carbonaceous grains when the distribution cuts off at size of $\sim 0.25~\mu$m. Grains smaller than 0.1$~\mu$m get trapped and destroyed in the remnant. However, the calculations by \citet{Slavin2020} have been constrained by early period (up to 8~kyr) of the SN adiabatic phase. More extended (up to 100~kyr) evolution has been considered by \citet{vs2024}, they came to almost the same conclusions as well. They found that small grains can later stick in the dense shell formed by the forward shock, when it becomes cold after the onset of radiative phase, thereby these grains survive against sputtering. Under these conditions the escaping efficiency of grains obviously decreases.

Large grains escaping from the SN remnant can travel through the ISM more than several thousand years with velocities higher than several hundreds km/s \citep{v26}. During this movement in the warm/cold ISM, grains can mainly suffer kinetic sputtering, while they have been decelerated below $300$~km/s, below which this process becomes negligible \citep{Nozawa2006}.  Here we focus on the mass of dust escaping from the SN remnant from the birth of grains in the SN ejecta to the age of several tens of thousand years, i.e. when the remnant becomes radiative. Section~2 describes the model and initial conditions. Section~3 presents the results. Section~4 summarizes the main conclusions.

%----------------------- Section 2 -------------------------------
\section{MODEL DESCRIPTION}\label{sec:mod}

We use a one-dimensional spherically symmetric Lagrangian approach to follow the evolution of a SN ejecta interaction with the ambient interstellar gas from early time, when the ejecta expands freely and dust begins to be produced inside it, $t_i \sim 100$~days \citep{Kozasa1989}, to the onset of the radiative phase in the SN shell, when the cooling time becomes close to the SN age: $t_c \sim t_d$ \citep[e.g.,][]{Wheeler1980,Shull1980}. For a gas with density of $1$~cm$^{-3}$ this timescale varies from $\sim 30$~kyr to $\sim 100$~kyr for metallicity equal ${\rm [Z/H]} = 0, \ -3$, respectively. The main parameters of our models can be found in Tab.~\ref{tab-models}.

%%%%%%%%%%%%%%%%%%%%%%%%%%%%%%%%%%%%%%%%%%%%%%%%%%%%%%%%%%%%%%
\begin{table}
\caption{The parameters of the models.}
\center
\begin{tabular}{lc}
\hline
\hline
{\bf The parameter}   & {\bf The value}  \\
\hline
\multicolumn{2}{c}{\it the SN ejecta}    \\
Energy &  $10^{51}$~erg  \\
Mass   &  $3~\msun$  \\
\hline
\multicolumn{2}{c}{\it the ISM}     \\
Density $n$ &  $1-100$~cm$^{-3}$  \\
Temperature $T$ &  $10^4$~K  \\
Metallicity ${\rm [Z/H]}$  &   0, $-3$ \\
\hline
\multicolumn{2}{c}{\it Dust}    \\
Maximum size $a_{cut}$  &  $0.3-10~\mu$m \\
Size range  &  $0.003~\mu$m -- $a_{cut}$  \\
Maximum number of sizes $N_s$ &  36    \\
\hline
\multicolumn{2}{c}{\it Grid}    \\
Number of cells per the ejecta &  110   \\
Total number of cells    &  $2000-10^4$  \\
\hline
\hline
\end{tabular}%
\label{tab-models}
\end{table}
%%%%%%%%%%%%%%%%%%%%%%%%%%%%%%%%%%%%%%%%%%%%%%%%%%%%%%%%%%%%%%

We consider the evolution of the SN ejecta with initial kinetic energy $E_k = 10^{51}$~erg, total mass $M_e = 3~\msun$. The maximum expansion velocity is given by $v_e = \sqrt{E_k/M_e}$. The density of the ambient gas is set to 1, 10 and 100~cm$^{-3}$. The gas temperature is equal to $10^4$~K.

For the initial distribution of gas density and velocity we adopt the radial profiles following from a self-similar solution for free expansion of the SN ejecta \citep{Truelove1999,ZeldRai}. In the density profile there are three regions: the core of the ejecta with the flat density distribution, the envelope of the ejecta with steep power-law decrease of density and the ambient gas. For the second we adopt the index $\alpha=9$. The two former expand freely with radially increasing velocity. The outer medium is at rest.  

We solve a 1D gas dynamical equations using the standard second-order accurate Lagrangian finite-difference scheme of constant mass zones \citep{Bowers1991,Thoul1995,Truelove1999}. We consider two regions: the SN ejecta and the ambient medium. The number of zones (numerical shells) in the ejecta is the same for all runs and equal to 110, which is sufficient for adequate modelling the dynamics of gas in the ejecta. To follow the evolution of SN ejecta with age from less than one year to thousands of years we set different spatial resolution in these two regions. The number of shells for the ambient region is varied for different models depending on the ambient gas density (the radius of the SN remnant is, obviously, larger in more rarefied gas) and the distance, which dust particles can attain (during their evolution dust grains should remain inside the grid). At the initial moment the sizes of the zones differ about a factor of several hundreds. Thus, the total number of zones in our runs remains reasonable and varies from 2000 to 10000 depending on the model. It is worth noting that we vary the factor from several to thousands to investigate a possible dependence on spatial resolution and find a good convergence for the value less than 2000. We also apply a logarithmically-increased grid and do not find any significant difference as well.  

At age of several tens of days we assume that dust is formed in a gas with $n\sim 10^8-10^{10}$~cm$^{-3}$, $T \simlt 10^4$~K \citep[e.g.][and many others]{Kozasa1989,Todini2001,Nozawa2006}. The formation process continues up to $\sim 10^3$~days \citep{Todini2001}. In our model we inject dust particles in the ejecta at the age $10^3$~days. We distribute the particles in the ejecta core along the radius uniformly. We add dust grains in the shells inside the core radius $r_c$ only, because the density in the ejecta envelope rapidly drops with radius. Initially we assume the velocity of particles is equal to that of the surrounding gas at radius $r$. Afterwards, at least during subsequent several tens of years, gas and grains are dynamically coupled. Thus, the dust dynamics in further times doesn't depend on the dust injection age, if this age is shorter the period, while the velocities of gas and grains have remained equal.

Since dust grains are distributed uniformly in the ejecta core and their velocities are equal to that of the ambient gas. Therefore, the grains move with the free expanding ejecta. Due to the interaction with gas they support their velocity equal to that of the surrounding gas flow, that is controlled by the stopping time scale \citep{Epstein1924,Baines1965}. We include the drag force acting on grains \citep{Draine1979b}, but, for simplicity, we neglect the backward reaction on the gas. The latter leads to that the dynamics of the gas both in the remnant and ambient medium is independent of the dust-to-gas ratio. The equations of motion for each particle are solved self-consistently with the 1D gas dynamics. We take into account the destruction of dust particles by both thermal (in a hot gas) and kinetic (due to a relative motion between gas and grains) sputtering \citep{Draine1979a}. For $T\sim 10^5 - 10^9$~K the typical timescale for thermal sputtering of silicate, graphite or iron grains can be estimated as \citep[eqn. 25.14 in][]{Draine-book}: 
\be
 t_{thr} \approx 10^5~ {\rm yr}~\left[ 1 + T_6^{-3}\right] {a_{0.1} \over n},
\label{eq:tsp}
\ee
where $T_6 = T/10^6$~K, $a_{0.1}$ is the grain size in units 0.1$~\mu$m, $n$ is the gas number density in units 1~cm$^{-3}$. The kinetic sputtering rate depends on the composition of grains \citep[see Fig.~2b in][]{Nozawa2006}, averaging over it we can use the approximation at $|v_p -v_{gas}|>50$~km~s$^{-1}$ \citep{vs2024}:
\be
 t_{kin} \approx 2.5\times 10^2~ {\rm yr} ~ {a_{0.1} \over n} ~ {\rm exp}\left({4\times 10^2 \over |v_p -v_{gas}|}\right) |v_p -v_{gas}|^{0.7},
\ee
where $v_{gas}$, $v_p$ are the velocities of gas and grains in km~s$^{-1}$.

It is important to know how mass of dust formed in the ejecta is transported over the remnant and further into the ISM. We apply the method of dust `superparticles' \citep[see for more details, e.g.][]{vs2024}. Each `superparticle' is supposed to represent an ensemble of numerous physical grains of a single size. To follow the dust transport we set one `superparticle'  per computational shell. Each `superparticle' contains the total mass of the ensemble of single-sized dust particles in it. 

We consider the dynamics of grains having a power-law Mathis-Rumpl-Nordsieck (MRN) size spectrum with index $-3.5$ \citep{Mathis1977}. Grain sizes are ranged from  $0.003~\mu$m to the cut-off (maximum) size $a_{cut}$, which is varied from 0.3 to 10~$\mu$m. At the initial moment (the time at which the grains have been added in the calculation), the number of grain sizes in the distribution is taken to be equal $N_{s}$, the initial sizes are binned equally in the log-scale. For the largest size $a_{cut}=10~\mu$m $N_s = 36$. Such approach allows us to compare the mass of dust escaping from the SN remnant for different cut-off values in the size distribution. Note that the upper limit in the distribution is much greater than the typical value of 0.1~$\mu$m and it can be caused by a possibility of the formation of grains with sizes larger than 1~$\mu$m in SN ejecta \citep[e.g.][]{Nozawa2003}. Grains of several microns can be considered for interpreting the spectral evolution of the bright SN \citep{Gall2014}. Even larger interstellar grains with size of $\sim 30~\mu$m have been detected in the Earth atmosphere \citep{Taylor1996,Baggaley2000,Baggaley2004}. Similar wide grain size distribution has been calculated inside the SN remnant after several hundreds of years evolution \citep{Kirchschlager2019}. Some discussion about the grain distribution at large sizes can be found in \citep[e.g.,][]{Draine2009}.

%----------------------- Section 3 -------------------------------
\section{RESULTS}\label{sec:res}

At early time the SN ejecta expands freely, therefore the forward shock sweeps up an ambient gas into the hot shell, while the reverse shock moves inside the cold dense ejecta and heats it \citep[e.g.,][]{Truelove1999}. When the shell becomes massive enough the SN remnant expands adiabatically \citep{Sedov1959}. The gas temperature in the shell goes down, when radiative losses in the shell are significant \citep{Oort1951}. The onset of the radiative phase comes with decelerating of the SN remnant expansion from $r\sim t^{2/5}$ to $\sim t^{2/7}$ \citep[e.g.,][]{Cioffi1988}. During this period the shell has become dense, and grains moving from the SN remnant interiors stick in the shell and cannot leave the remnant \citep[e.g.,][]{vs2024,Dedikov2025na}. Thus, we get the mass fraction of dust produced in the SN ejecta and escaping from the remnant till this age.

%%%%%%%%%%%%%%%%%%%%%%%%%%%%%%%%%%%%%%%%%%%%%%%%%%%%%%%%%%%%%%%%%%%%%
\begin{figure*}
\center
\includegraphics[width=9.5cm]{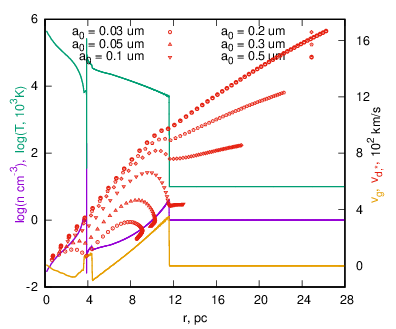}
\caption{
The radial profiles of gas density, temperature (left axis) and velocities (right axis) of gas (shown by line) and dust grains (shown by symbols) at $t=10$~kyr for the SN remnant expanding in the ambient gas with density 1~cm$^{-3}$. The symbols depict the velocity of grains with the initial size 0.03, 0.05, 0.1, 0.2, 0.3 and 0.5~$\mu$m (see the legend). 
}
\label{fig-pro-ev}
\end{figure*}
%%%%%%%%%%%%%%%%%%%%%%%%%%%%%%%%%%%%%%%%%%%%%%%%%%%%%%%%%%%%%%%%%%%%%

Figure~\ref{fig-pro-ev} presents the radial profiles of gas density, temperature and velocity of gas at $t=10$~kyr for the SN remnant expanding in the ambient gas with density $n=1$~cm$^{-3}$. At this time moment the remnant has a typical structure for adiabatic phase \citep[e.g.,][]{Cioffi1988}: the contact discontinuity at the radial distance $r\sim 4$~pc separates the ejecta shocked by the reverse shock from the swept-up interstellar gas heated by the forward shock located at $r\sim 12$~pc. 

The dynamics of dust grains is determined by their sizes and the evolution of a gas. Their dynamics is described in detail in \citep[e.g.,][]{Slavin2020,v26}. Here the main points are briefly reported. Figure~\ref{fig-pro-ev}  shows the velocities of dust grains for several sizes. Initially (at the dust formation time moment), the velocity of grains is assumed to be equal to that of the surrounding gas. Grains move with the free expanding ejecta. Due to both interaction with the reverse shock and growth of the swept-up shell, grains enter into a hot gas behind the shocks, suffer sputtering and start to decouple from the slowing-down  flow of the swept-up shell. 

Small grains are less inertial and easily follow the gas flow in the remnant. Grains being initially smaller $a_0\simlt 0.05~\mu$m remain inside the remnant within the whole their evolution. They are efficiently destroyed in the hot bath of the remnant. Dust particles larger $a_0 \simgt 0.1~\mu$m born in the outer core of the ejecta keep higher velocity, move through the hot swept-up shell, overtake the forward shock front and leave the remnant \citep{Slavin2020,vs2024,v26}. Due to higher velocity they spend shorter time in hot gas of the remnant and, therefore, they are destroyed by sputtering less. Afterwards, during their moving in the ISM with high velocity, they attain the distances exceeding the remnant's radius substantially \citep{v26}.

%%%%%%%%%%%%%%%%%%%%%%%%%%%%%%%%%%%%%%%%%%%%%%%%%%%%%%%%%%%%%%%%%%%%%
\begin{figure*}
\center
\includegraphics[width=9.5cm]{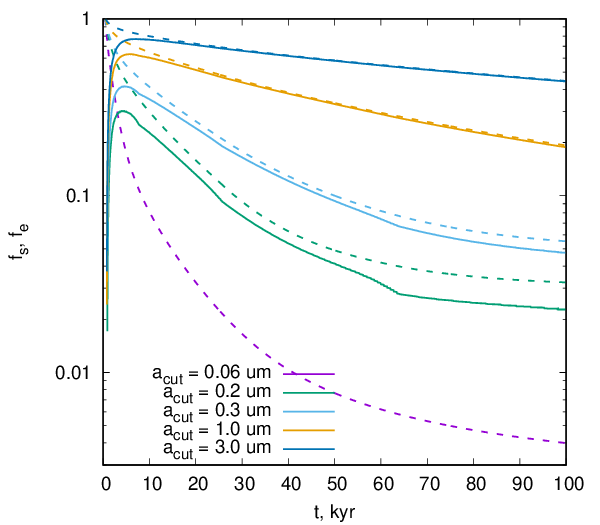}
\caption{
The mass fraction of the surviving (dash lines) and escaping (solid line) dust from the SN remnant expanding in the ambient gas with density 1~cm$^{-3}$. The lines of different color correspond to the cut-off size $a_{cut}$ in a MRN distribution of grains: 0.06, 0.2, 0.3, 1 and 3~$\mu$m. The absence of the solid line for the cut-off size of 0.06~$\mu$m means there is no grains escaping from the remnant: all dust produced in the ejecta remains confined inside the remnant's radius.
}
\label{fig-mdc-out-evol}
\end{figure*}
%%%%%%%%%%%%%%%%%%%%%%%%%%%%%%%%%%%%%%%%%%%%%%%%%%%%%%%%%%%%%%%%%%%%%

%%%%%%%%%%%%%%%%%%%%%%%%%%%%%%%%%%%%%%%%%%%%%%%%%%%%%%%%%%%%%%%%%%%%%
\begin{figure*}
\center
\includegraphics[width=9.5cm]{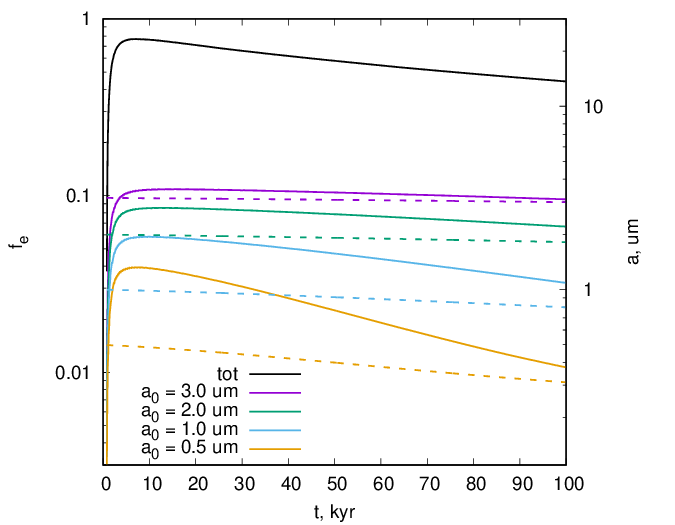}
\caption{
The total mass fraction of the dust escaping from the SN remnant expanding in the ambient gas with density 1~cm$^{-3}$ (black solid line) for the cut-off size $a_{cut}=3~\mu$m in a MRN distribution of grains. The contributions to the total fraction from grains with size $a_0$ are depicted by colored solid lines. The evolution of grains with initial size $a_0$ is shown by colored dashed lines on the right axis.
}
\label{fig-mdc-out-evol-frac31}
\end{figure*}
%%%%%%%%%%%%%%%%%%%%%%%%%%%%%%%%%%%%%%%%%%%%%%%%%%%%%%%%%%%%%%%%%%%%%

%%%%%%%%%%%%%%%%%%%%%%%%%%%%%%%%%%%%%%%%%%%%%%%%%%%%%%%%%%%%%%%%%%%%%
\begin{figure*}
\center
\includegraphics[width=9.5cm]{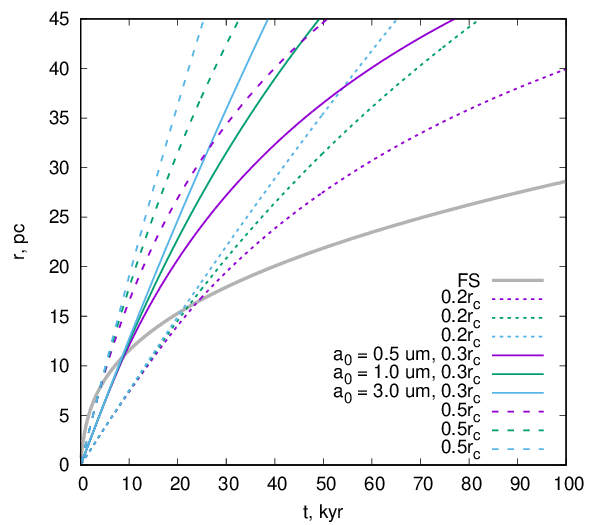}
\caption{
The radial position of a grain initially located at distance $0.2$, $0.3$ and $0.5~r_c$ (shown by dot, solid and dash lines, respectively). The tracks for a grain with initial size $a_0$ equal to $0.5$, $1$ and $3~\mu$m are depicted by lines of different color. The grey line shows the SN remnant radius, i.e. the position of the forward shock.
}
\label{fig-rd-evol1}
\end{figure*}
%%%%%%%%%%%%%%%%%%%%%%%%%%%%%%%%%%%%%%%%%%%%%%%%%%%%%%%%%%%%%%%%%%%%%

Thus, we can determine two values: at first, the mass fraction of dust surviving in the SN remnant $f_s$, which includes all dust produced in the SN ejecta and then spreading over both remnant and ambient gas, and, at second, the mass fraction of dust escaping from the SN remnant and moving in the ISM $f_e$. It is interesting to consider, how the variation of the cut-off size $a_{cut}$ of grains from 0.06 to 3~$\mu$m affects on these values. Figure~\ref{fig-mdc-out-evol} presents the evolution of the mass fraction of the dust surviving in the SN remnant expanding in the ambient gas with density $n=1$~cm$^{-3}$. Grains smaller 0.06~$\mu$m are destroyed most efficiently: the dust mass fraction decreases about a factor of 100 by $t\sim 30$~kyr and about 300 times by the age of $\sim 100$~kyr. Increase of the cut-off size $a_{cut}$ leads to growth of the surviving fraction. One can see that for the SN remnant age greater than 20~kyr the surviving fraction of increases about an order of magnitude for $a_{cut} = 0.2~\mu$m and almost two orders of magnitude for $a_{cut} = 1~\mu$m (green and yellow dotted lines, respectively).
 
In what follows, we compare our estimates of the surviving fraction of grains with previous results. It is apparent from Figure~\ref{fig-mdc-out-evol} that by the age $\approx 8$~kyr the mass fraction of surviving dust is around  $35-40$\% for the grains cut-off size $0.2-0.3~\mu$m in the distribution shown by green and blue dash line, respectively. This value appears to be close to the estimate obtained by \citet{Slavin2020} for graphites with initial power-law MRN distribution cutted off at $0.25~\mu$m. \citet{vs2024} adopted a monodispersed initial distribution of grains: for the initial size of $0.3~\mu$m the mass fraction drops to $\sim 0.2$ by the age of 100~kyr for the SN remnant expanding in a gas with density $1$~cm$^{-3}$ \citep[see Figure~5 im][]{vs2024}. Figure~\ref{fig-mdc-out-evol} shows that the fraction of surviving dust is about $0.06$ for the same cut-off grain size and SN age. However, \citet{vs2024} assumed that all dust initially had a single size, while in the model considered here the size distribution follows a power-law: the initial mass fraction pf large ($\sim 0.1-0.3~\mu$m) grains surviving in the remnant by 100~kyr. is $\approx 0.5$, while the fraction of surviving dust estimated here equals to $\sim 0.06/0.5 \approx 0.12$, which is close to the value obtained by \citet{vs2024}.

A slope of the surviving fraction depends on the cut-off size of grains (Figure~\ref{fig-mdc-out-evol}). It becomes flatter for larger cut-off value. Moreover, for smaller cut-off size it is also flatter at late age. This behaviour is determined by both destruction efficiency of grains and their dynamics. The surviving fraction is decreased in time for any cut-off value in a power-law distribution. This will be continued until the cessation of dust destruction both inside the remnant due to reaching radiative phase and outside the remnant due to decelerating grains down to $\sim 300$~km/s. Afterwards grains continue slowing-down, but are not destroyed neither thermal nor kinetic sputtering.

Figure~\ref{fig-mdc-out-evol} presents the evolution of the mass fraction of the escaping dust from the SN remnant. After a rapid growth during several first kyrs of the evolution it follows to the surviving fraction. For larger cut-off size $a_{cut}$ these fractions become closer to each other, i.e. all dust surviving during the SN remnant's evolution escapes the remnant. For the cut-off size $a_{cut} = 3~\mu$m almost 70\% of the dust mass produced by SN ejecta has left the remnant by the age of 30~kyr, when the shell expanding in the solar metallicity ambient gas becomes radiative. For lower metallicity the shell remains adiabatic for longer ages. Grains continue escaping the remnant, however, they are sputtered due their high velocity. Therefore, in a gas with ${\rm [Z/H]} \sim -3$ only 45\% of the initial dust mass produced by SN ejecta survives outside of the remnant. 

Note that grains of various size make different deposits to the escaping fraction proportional to their destroying efficiency and depending on their dynamics. Figure~\ref{fig-mdc-out-evol-frac31} presents the contributions to the escaping fraction from grains with the initial size $a_0$ for the cut-off size $a_{cut}=3~\mu$m. As one remember that large grains dominate by mass in a MRN distribution. For instance, about 10\% by mass is deposited by grains\footnote{Note that for the cut-off size $a_{cut}=3~\mu$m the distribution from $0.003~\mu$m to $a_{cut}$ at the initial moment (the moment when grains have been added into calculations)  contains $N_{s}=31$ sizes equally allocated in the log-scale.} with the initial size $a_0=3~\mu$m (see solid magenta line and left axis in Figure~\ref{fig-mdc-out-evol-frac31}). Grains with $a_0=3~\mu$m are destroyed very marginally (see dash magenta line and right axis in Figure~\ref{fig-mdc-out-evol-frac31}). Therefore, the contribution from these grains remains almost constant during the evolution. Grains with $a_0=1~\mu$m are decreased in size to $\sim 0.8~\mu$m or by a factor of $\sim 1.25$ (light-blue dash line, right axis). This is manifested in decreasing the escaping fraction of such grains by a factor of two (light-blue solid line).

According to our model grains are uniformly allocated in the ejecta core and, therefore, they get their initial velocity proportional to their radial position in the core: outer particles have higher velocity than inner ones. Owing to that larger grains keep longer their velocities close to the initial value, such grains escape the remnant earlier. Initially about 50\% of the dust mass produced in the SN ejecta contains in grains larger $a_0\geq 1~\mu$m for the power-law distribution with the cut-off size $a_{cut}=3~\mu$m. Figure~\ref{fig-rd-evol1} shows the radial tracks of large dust particles. Grains of size larger than $1~\mu$m  initially located at distance $\simgt 0.2~r_c$ have overtaken the forward shock by the age a little over $\sim 20$~kyr. Afterwards they move in the ambient medium and keep their size close to the initial value. Such large grains represent the vast fraction: $[M_d(r<r_c) - M_d(r>0.2r_c)]/M_d(r<r_c) = (1-0.2^3)\approx 0.99$, where $M_d(r<r_c)$ is the mass of the dust inside radius $r_c$ at time of its formation, i.e. more than 99\%, of the dust mass produced in the ejecta core. Therefore, grains with $a_0 \geq 1~\mu$m for $a_{cut}=3~\mu$m transport about 50\% of the dust mass from the remnant to the ambient medium by the age about 20~kyr for the SN remnant expanding in a gas with density 1~см$^{-3}$.

%%%%%%%%%%%%%%%%%%%%%%%%%%%%%%%%%%%%%%%%%%%%%%%%%%%%%%%%%%%%%%%%%%%%%
\begin{figure*}
\center
\includegraphics[width=9.5cm]{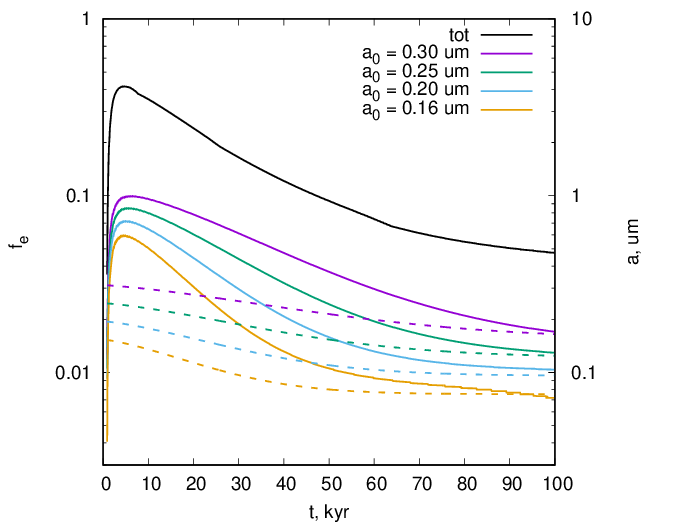}
\caption{
The same as in Figure~\ref{fig-mdc-out-evol-frac31}, but for the cut-off size $a_{cut}=0.3~\mu$m.
}
\label{fig-mdc-out-evol-frac21}
\end{figure*}
%%%%%%%%%%%%%%%%%%%%%%%%%%%%%%%%%%%%%%%%%%%%%%%%%%%%%%%%%%%%%%%%%%%%%

%%%%%%%%%%%%%%%%%%%%%%%%%%%%%%%%%%%%%%%%%%%%%%%%%%%%%%%%%%%%%%%%%%%%%
\begin{figure*}
\center
\includegraphics[width=9.5cm]{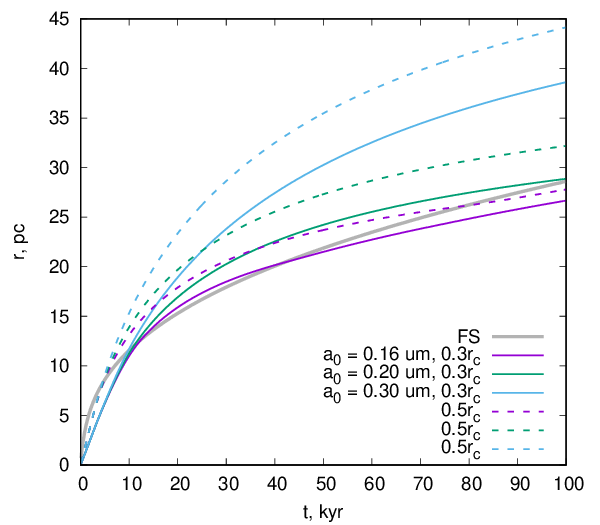}
\caption{
The radial position of a grain initially located at distance $0.3$ and $0.5~r_c$ (shown by solid and dash lines, respectively). The tracks for a grain with initial size $a_0$ equal to $0.16$, $0.2$ and $0.3~\mu$m are depicted by lines of different color. The grey line shows the SN remnant radius, i.e. the position of the forward shock.
}
\label{fig-rd-evol}
\end{figure*}
%%%%%%%%%%%%%%%%%%%%%%%%%%%%%%%%%%%%%%%%%%%%%%%%%%%%%%%%%%%%%%%%%%%%%

%%%%%%%%%%%%%%%%%%%%%%%%%%%%%%%%%%%%%%%%%%%%%%%%%%%%%%%%%%%%%%%%%%%%%
\begin{figure*}
\center
\includegraphics[width=9.5cm]{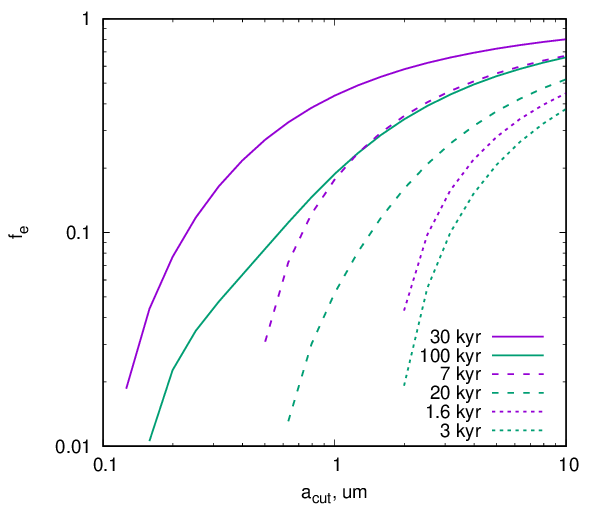}
\caption{
The mass fraction of the dust escaping from the SN remnant vs the cut-off size $a_{cut}$ in a MRN distribution of grain sizes. The lines of different type show the dependence for the remnant evolving in the ambient gas with density $n = 1$, 10 and 100~cm$^{-3}$ (solid, dash and dot lines respectively). The color of lines depicts the fraction at the age of the SN remnant corresponding to the onset of radiative phase in the remnant at metallicity of an ambient gas: ${\rm [Z/H]}=0$ (upper line) and ${\rm [Z/H]}=-3$ (lower line). The number in the legend shows this age in kyr.
}
\label{fig-admx-mdc}
\end{figure*}
%%%%%%%%%%%%%%%%%%%%%%%%%%%%%%%%%%%%%%%%%%%%%%%%%%%%%%%%%%%%%%%%%%%%%

In case of smaller cut-off size the escaping fraction following to the surviving one changes the slope during the evolution. Figure~\ref{fig-mdc-out-evol-frac21} shows the contributions for the cut-off size $a_{cut}=3~\mu$m. After reaching the maximum at $\sim 3-5$~kyr the escaping fraction of grains with various sizes decreases with different slope: it is steeper for smaller grains. But later the dependences have the same slope. Small grains spend some period in the hot gas of the remnant, where they are efficiently sputtered. Some of them can overtake the forward shock at early age of the remnant and move through an ambient gas. However, they can be decelerated there, so that their velocities become lower than that of the incoming forward shock. Therefore, in some conditions these grains can enter in the hot gas once again, in which their velocities do not decrease. In time the forward shock decelerates as well, so that some grains can overtake it again. Figure~\ref{fig-rd-evol} shows the radial tracks of some grains. One can see that grains of size $0.16~\mu$m initially located at distance $0.3~r_c$ (solid magenta line) are inside the hot remnant before $\sim 10$~kyr, afterwards they have overtaken the forward shock and travelled in an ambient gas until $\sim 40$~kyr, when they have crossed the shock and enter in the hot gas again. This behaviour enhances the destruction efficiency of such grains, that leads to decreasing the total surviving and escaping fractions. One should note that radiative losses become significant at this age, that leads to slowing down the forward shock expansion, therefore it cannot catch up grains being ahead the front (see, e.g. solid magenta line in Figure~\ref{fig-rd-evol}). Moreover, the shell being radiative is becoming denser and grains located behind it and moving outside can stick in it.

Figure~\ref{fig-admx-mdc} presents the mass fraction of the dust escaping from the SN remnant evolving in an ambient gas with density $n = 1$, 10 and 100~cm$^{-3}$ for the wide range of the cut-off size $a_{cut}$ in a MRN distribution. The upper lines correspond to the SN ages, when  radiative phase begins in the remnant evolving in a gas with metallicity ${\rm [Z/H]}=0$ and the lower lines depict the fraction for a gas with ${\rm [Z/H]}=-3$ (see the labels in the legend). Owing to the escaping fraction follows closely the surviving one at late age of the SN remnant, we get estimates for both values. One can note that these fractions grow with increase of the cut-off value in the distribution of grain sizes. For large cut-off value $a_{cut}$ the fractions tend to be saturated for a given age of the SN remnant. For example, the fractions reach about 0.8 at age of 30~kyr for the largest cut-off size of $10~\mu$m. That means that almost 80\% of the mass of dust produced in the SN ejecta has escaped out from the remnant to the ISM by this age. This timescale corresponds to the onset of radiative phase in a solar metallicity gas. Afterwards grains cannot leave the remnant owing to their sticking in the radiative dense shell. However, almost 99\% (by mass) of grains larger than $1~\mu$m have left the remnant by the age of 30~kyr.

Such grains travel in the ISM and suffer kinetic sputtering. Therefore, the fraction will gradually decrease later because of high-velocity large grains. One can see that by the age of 100~kyr grains of $a_0\simgt 2~\mu$m decrease in size by less than 4\% (green dash line, see right axis of Figure~\ref{fig-mdc-out-evol-frac31}). While their velocity remains sufficiently high as $\sim 400-600$~km/s, they continue to be destroyed. In the next 100~kyr their velocity drops below $\sim 300$~km/s, when kinetic sputtering becomes inefficient \citep{Nozawa2006}, during this period the size of grains of $a_0\simgt 2~\mu$m will reduce another $\sim 5$\%. That means that the dust mass diminishes less than 30\%, for $a_{cut} = 10~\mu$m the escaping/surviving fraction for the ambient gas density $n\sim 1$~cm$^{-3}$ goes down from almost 70\% at 100~kyr (green solid line in Figure~\ref{fig-admx-mdc}) to $\sim 50$\% at $\sim 200$~kyr.

In denser environment the escaping fraction decreases significantly for $a_{cut} \simlt 1-3~\mu$m. For the ambient gas with density $n\sim 100$~cm$^{-3}$ the fraction retains higher than 0.1 for $a_{cut} \simgt 3~\mu$m (see dot lines in Figure~\ref{fig-admx-mdc}). The onset of the radiative phase in such a dense gas takes place at much shorter age. The dynamics of grains has the same features as in the low-dense gas, which are described above, but this is valid for larger grain sizes. For instance, grains of $a_0\sim 1-2~\mu$m are destroyed moderately in a gas with $n\sim 1$~cm$^{-3}$, while they are completely destroyed in the medium with gas density $\sim 100$~cm$^{-3}$. In such dense medium only very large grains in the distribution with $a_{cut} \simgt 5~\mu$m are able to escape from the remnant and afterwards survive at a significant amount. These grains propagate at several parsecs from the forward shock front only. Such distance is comparable to the sizes of dense molecular clouds, in which SN can explode. Therefore, dust with the initial size $a_{cut} \simgt 5~\mu$m produced by SN remains inside its parent molecular clouds, while in low-dense diffuse medium such grains can spread over the volume of more than $100$~pc in radius.

One should note that in the presence of possible extra cooling from grains in hot gas \citep{Burke1974,Ostriker1973,Silk1974,Smith1996,Seok2013} the remnant expansion is slowed down earlier \citep{Dedikov2025irx}, therefore, the escaping fractions might be slightly higher. 

Above we consider a widely-used power-law distribution of grain sizes with index $-3.5$ \citep{Mathis1977}, while there are more observationally motivated distribution functions \citep{Jones1996,Weingartner2001,Hensley2023}. In some of these distributions there is an excess of small grains, which cannot escape hot SN bubble as it has been shown above. However, these distributions are for dust grains processed by the reverse shock and numerous interstellar shocks. Actually, the reverse shock \citep{Bianchi2007,Nozawa2007} considerably reduces the fraction of small size dust particles \citep[$a\simlt 10^{-3}\mu$m, see][lower panel in their Fig. 6]{Nozawa2007}. While large grains containing the major mass fraction of dust produced by SN can survive and escape the parent SN bubble with high velocity. Several studies are favour to a ``flat'' spectrum of grain sizes $n_d(a) \propto a^{-1.5}$ \citep{Maiolino2004,Nozawa2007} for conditions in early galaxies. In such case the mass fraction of dust escaping SN bubble is expected to be higher, this issue will be studied in details elsewhere.

It is known that magnetic fields in the ISM can substantially affect the dynamics of dust particles. A charge of small grains with size of $\sim 0.02~\mu$m varies rapidly and becomes close to zero \citep{Weingartner2001,Ibanez2019}. However, grains of size up to $0.1~\mu$m cannot leave the SN remnant. On the other hand, grains larger $0.1~\mu$m can penetrate into the ISM and might be tightly coupled with magnetic field. Subsequently, they can be either accelerated \citep{Lazarian2002,Yan2004}  or slowed down \citep{McKee1987}. However, for such large dust particles the influence of grain charge on their dynamics will be suppressed because  the grains charge-to-mass ratio decreases with increasing size as $q/m \sim a^{-2}$ \citep[e.g.,][]{Slavin2012}. Moreover, the dynamics of dust particles depends on the magnetic field structure and requires more detailed study in three dimensional calculations.

%----------------------- Section 4 -------------------------------
%\section{DISCUSSION}\label{sec:dis}

%...

%----------------------- Section 5 -------------------------------
\section{CONCLUSIONS}\label{sec:con}

Here we have considered the efficiency of survival and escape from the SN remnant for dust grains formed in the SN ejecta from free expansion to late adiabatic phase until the age, when the SN shell becomes radiative. We estimate the surviving and escaping mass fractions depending on the maximum (cut-off) size in a power-law distribution of grain sizes with index $-3.5$. The results are summarized as follows:
\begin{itemize}
 \item the mass fractions of the surviving and escaping dust from the SN remnant grow for larger cut-off size in a power-law distribution with index $-3.5$; for the cut-off size $a_{cut} = 1~\mu$m almost 45\% of the dust mass produced by SN ejecta can leave the remnant expanding in a gas with density 1~cm$^{-3}$ by the age of 30~kyr; the fraction increases to about 70\% for $a_{cut} = 3~\mu$m;

 \item grains of size larger than $1~\mu$m  initially located in the ejecta core at radial distances $\simgt 0.2~r_c$ overtake the forward shock of the SN remnant evolving in a gas with density 1~cm$^{-3}$ by $\sim 30$~kyr; for $a_{cut} = 3~\mu$m such large grains represent more than 99\% of the dust mass produced in the ejecta core and transport a little less than 50\% of the dust mass from the remnant to the ambient medium of density 1~cm$^{-3}$ by $\sim 30$~kyr;

 \item high-velocity dust grains escaping the remnant continue to be destroyed further due to kinetic sputtering in the interstellar medium; grains with sizes of several microns are destroyed only marginally and stopped within timescale about one hundred kyr for remnant expanding in a gas with density 1~cm$^{-3}$; 
 
 \item the surviving fraction of dust produced in the SN remnant, which expands in the ambient gas with $n$ from 1 to 100~cm$^{-3}$ varies from $\sim 0.5$ to $\sim 0.3$, respectively, for the cut-off size $a_{cut} = 10~\mu$m; a decrease of the cut-off value in the distribution reveals in dropping of the surviving mass fraction of dust.
\end{itemize}

It is apparent from the results, that the dynamics of dust grains depends significantly on their birthplace in the ejecta: grains formed at large radial distances have higher velocities. Thus, larger grains leave the SN remnant easier, and the mass of dust surviving and entering in the ambient interstellar medium increases. The 'lack of dust' problem is probably to be slightly weakened owing to the transport of a significant mass fraction of large grains produced in the ejecta outside of the SN remnant \citep[see the discussion in][]{Mattsson2021,Kirchschlager2022,Peroux2023}. Another consequence could be a more efficient outflow of dust from galaxies \citep{Ferrara1990,Shustov1995,Shchekinov2011,Shchekinov2025rev}, since large grains can move with velocities up to one thousand km/s and attain heights up to several hundreds parsecs above the galactic midplane within timescale shorter than one million years. In this case grain velocities are higher than the escape velocity even for massive galaxies. Later such grains can pass into outer ($r \simgt 100$~kpc) halo and leave the parent galaxy, and forming an extended dusty halo \citep[e.g.,][]{Ferrara1994,Menard2010,Yoon2021}.

%----------------------- Section A -------------------------------
\section*{Acknowledgements}

The author is grateful to the referee for the numerous suggestions that significantly improved the paper. The author is thankful to Yuri Shchekinov and Svyatoslav Dedikov for many discussions and Ilya Khrykin for his invaluable assistance.

%\section*{FUNDING}

\section*{CONFLICT OF INTEREST}

The authors of this work declare that they have no conflicts of interest.

%----------------------- Section L -------------------------------
\bibliographystyle{aspb1}
\bibliography{p-bib1.bib}

\end{document}